\documentclass[aps,pra,twocolumn,showpacs,floatfix]{revtex4}

\usepackage{epsf}
\usepackage{amsmath}
\usepackage{amsfonts}
\usepackage{amssymb}
\usepackage{bm}
\usepackage{bbm}
\usepackage{nicefrac}
\usepackage{color}
\usepackage{pifont}
\usepackage{graphicx}
\usepackage{dcolumn}

\newcolumntype{.}{D{x}{}{-1}}

\def\ii{{\mathrm{i}}}

\newcommand{\Za}{\mbox{$Z\,\alpha$}}

\begin{document}

\title{QED corrections of order $\bm{\alpha(Z\alpha)^2E_F}$ to the hyperfine splitting\\
of $\bm{P_{1/2}}$ and $\bm{P_{3/2}}$ states in hydrogenlike ions}

\author{Ulrich D. Jentschura}
\affiliation{Department of Physics,
Missouri University of Science and Technology,
Rolla, Missouri 65409-0640, USA}
\affiliation{Institut f\"{u}r Theoretische Physik,
Universit\"{a}t Heidelberg,
Philosophenweg 16, 69120 Heidelberg, Germany}

\author{Vladimir A. Yerokhin}
\affiliation{Max--Planck--Institut f\"ur Kernphysik,
Postfach 10 39 80, 69029 Heidelberg, Germany}
\affiliation{Center
for Advanced Studies, St.~Petersburg State Polytechnical
University, Polytekhnicheskaya 29, St.~Petersburg 195251, Russia}

\begin{abstract}
The hyperfine structure (HFS) of a bound electron is modified by the
self-interaction of the electron with its own radiation field. This effect is
known as the self-energy correction. In this work, we discuss the evaluation of
higher-order self-energy corrections to the HFS of bound $P$ states.  These are
expressed in a semi-analytic expansion involving powers of $Z\alpha$ and
$\ln(Z\alpha)$, where $Z$ is the nuclear charge number and $\alpha$ is the
fine-structure constant. We find that the correction of relative order $\alpha
\, (Z\alpha)^2$ involves only a single logarithm $\ln(Z\alpha)$ for $P_{1/2}$
states [but no term of order $\alpha \, (Z\alpha)^2 \ln^2(Z\alpha)$], whereas
for $P_{3/2}$ states, even the single logarithm vanishes.  By a
Foldy--Wouthuysen transformation, we identify a nuclear-spin dependent
correction to the electron's transition current, which contributes to the HFS
of $P$ states.  A comparison of the obtained analytic results to a numerical
approach is made.
\end{abstract}


\maketitle


\newpage

%
%
\section{Introduction}

Throughout the history of quantum electrodynamic (QED)
calculations of atomic properties (notably, energy shifts),
two approaches have mutually inspired each other, namely,
the analytic and the numerical {\em ansatz}. The necessity for employing 
both methods
is easily seen when one considers the range of coupling constants
$Z\alpha$ which are relevant for hydrogenlike ions (here,
$Z$ is the nuclear charge number, and $\alpha$ is the fine-structure
constant). For low $Z$, the parameter $Z\alpha$ is very small,
and an expansion of the fermionic propagators in powers 
of $Z\alpha$ appears reasonable. For high $Z$, the parameter $Z\alpha$ 
approaches unity, and the mentioned expansion is not practical~\cite{MoPlSo1998}.
Consequently, starting with the accurate investigations on the 
spectrum of highly charged ions~\cite{Mo1974a,Mo1974b}, 
there has been tremendous activity over the past few decades 
regarding the accurate description of energy levels of heavy ions.
These are complemented by technically demanding 
experiments~\cite{GuEtAl2005}. 

Some of the most important effects to be 
considered in QED bound-state calculations are so-called 
self-energy corrections, where a bound electron
spontaneously emits and reabsorbs a virtual photon.
Between the photon emission and absorption, other 
interactions---with the binding Coulomb field 
and, possibly, with other external fields---may occur.
Even for low-$Z$ ions, a direct expansion of the 
electron propagators 
in powers of $Z\alpha$ is not universally possible: namely, 
the energy of the virtual photon must be large enough (larger than the 
scale of the atomic binding), so that the electron can essentially 
be regarded as a free particle in between the emission and absorption
(perturbed only by a finite number of interactions with the 
Coulomb field). However, when the photon energy 
is small (commensurate with the atomic binding energy), this expansion
is no longer possible. In this case, the expansion in powers
of $Z\alpha$ is achieved ``implicitly,'' by observing that 
since the photon energy is so small, one may expand the 
currents at the photon emission and absorption vertices in the 
long-wavelength limit, i.e.~in terms of dipole interactions, 
quadrupole interactions, spin-dependent interactions 
etc.~(see Ref.~\cite{Pa1993,JePa1996}).
The Foldy--Wouthuysen transformation~\cite{FoWu1950} can be used
in order to achieve a clear separation of the Hamiltonian 
into a leading nonrelativistic term and relativistic corrections,
and a decoupling of upper and lower components of the 
Dirac wave function to a specified order in $Z\alpha$
is achieved. This transformation, together with a careful
matching procedure needed in order to ``join'' the high-and low-energy 
parts, then leads to the analytic results traditionally used
in order to describe the 
Lamb shift~\cite{ErYe1965ab,Pa1993} and other effects
such as the bound-electron $g$ factor~\cite{PaCzJeYe2005}.

The intricacies described above are responsible for the
$Z\alpha$ expansion not being a simple power expansion (Taylor series)
in $Z\alpha$. The matching of high-and low-energy contributions at 
an intermediate photon energy scale commensurate with 
an overlapping parameter $\epsilon$ 
leads to the appearance of logarithms of $Z\alpha$
(see also the illustrating example in Appendix~A of~\cite{JePa2002}).
The expansion thus is semi-analytic.
For the Lamb shift in hydrogenlike systems, 
many terms have been calculated in the semi-analytic expansion~\cite{Pa1993}, 
but the numerical approach was faced with tremendous 
problems, and no direct result had been calculated 
for hydrogen ($Z =1$) up to the year 1999.
At low $Z$, the experimental accuracy is orders of magnitude higher
than at high $Z$, and the comparison and the implied 
mutual consistency check of the numerical and 
analytic calculations is most meaningful. 
Therefore, a calculation was carried out~\cite{JeMoSo1999}
which confirmed the consistency of both approaches
and determined a nonperturbative remainder term
which is beyond the sum of the known 
terms in the $Z\alpha$ expansion. The nonperturbative remainder 
term amounts to roughly 28~kHz for the hydrogen ground state Lamb shift,
which is a numerically large effect as compared to the 
current experimental accuracy of about 
33~Hz for the $1S$--$2S$ transition~\cite{NiEtAl2000,FiEtAl2004}.
Similar calculations were done, and agreement of the analytic and 
numerical calculations was found, for $P$ 
states~\cite{JeMoSo2001pra}. Other effects studied both within the 
$Z\alpha$ expansion and within the numerical approach include the 
bound-electron $g$ factor of a bound $S$ 
state~\cite{YeInSh2004,PaCzJeYe2005}. 
Also,
the self-energy correction to the hyperfine structure (HFS) was extensively
studied for the $S$ states during last decades, both within the numerical
all-order approach  \cite{BlChSa1997reserve,SuPeSaScLiSo1998,YeSh2001hyp}
and within the $Z\alpha$ expansion \cite{Pa1996,NiKi1997}. 

For the hyperfine splitting of $P$ states, however, investigations of the 
self-energy corrections are much more scarce, both within the numerical 
as well as within the analytic approach. There are some
quite recent all-order numerical calculations for the $2P_{1/2}$ state
\cite{SaCh2006} as well as for the $2P_{3/2}$ level~\cite{SaCh2008} (see also the
latest paper~\cite{YeJe2009}), but there are only few analytic results
to compare to. If we denote the nonrelativistic Fermi energy by 
$E_F$, then the only known self-energy correction~\cite{BrPa1968}
is that of order $\frac{\alpha}{\pi} E_F$.
This correction amounts to  
$\frac{\alpha}{4 \pi} E_F$ for $nP_{1/2}$ states and
$-\frac{\alpha}{8 \pi} E_F$ for $nP_{3/2}$ states.
Because this correction is entirely 
due to the electron magnetic moment,
we can refer to it as an anomalous magnetic moment correction 
to the HFS.

In this work, in the order to address the current, somewhat unsatisfactory
status of theory, we calculate the self-energy correction to the HFS of $P$
states up to order $\alpha (Z\alpha)^2E_F$.  This correction goes beyond the
anomalous magnetic moment correction and is due to numerous effects.  Indeed,
the calculation constitutes quite a complicated problem, mainly for three
reasons. First, the problem considered is a radiative correction under the
influence of an additional external field, i.e., we have three fields to
consider (the photon field, the Coulomb field and the nuclear magnetic field).
Second, the angular algebra is much more complicated than for reference $S$
states~\cite{JeYe2006}. The third difficulty is that for $P$ states, the
nonrelativistic limit of the hyperfine interaction has to be treated much more
carefully than for $S$ states. Namely, we may anticipate that there exists a
correction to the electron's transition current, caused by the hyperfine
interaction, which contributes to the self-energy correction for $P$ states,
but vanishes after angular integration for $S$ states. This correction to the
current can be understood easily if one considers the coupling of the physical
momentum $\vec{p} - e \, \vec{A}_{\rm hfs}$ of the electron to the vector
potential of the quantized electromagnetic field [here, $\vec{A}_{\rm hfs}$ is
the vector potential corresponding to the hyperfine interaction defined below
in Eq.~\eqref{Ahfs}].  The correction to the current has not been considered in
the previous investigations which dealt with $S$
states~\cite{Pa1996,NiKi1997,JeYe2006}.

We organize our investigation as follows.
In Sec.~\ref{general}, we present some known formulas needed for the 
description of the HFS. 
In Sec.~\ref{fw}, we consider the Foldy-Wouthuysen transformation
of the Hamiltonian, and of the transition current, and 
identify all terms relevant for the current investigation.
In Sec.~\ref{lterm}, the logarithmic 
terms of order $\alpha (Z \alpha)^2 \ln(Z\alpha) E_F$ are given special 
attention. Their value is derived within a straightforward 
and concise analytic approach.
We then continue, in Sec.~\ref{hep}, to investigate the 
contribution of high-energy photons to the HFS of 
$P$ states. Six effective operators are derived which are 
evaluated for general principal quantum number of the reference
state. The low-energy part is treated next (see Sec.~\ref{lep}).
The vacuum-polarization correction is obtained in Sec.~\ref{sec:vp}. 
The results are summarized in Sec.~\ref{final}
and conclusions are drawn in Sec.~\ref{conclu}.
Natural units ($\hbar = c = \epsilon_0 = 1$) are 
used throughout this paper.

%
%
\section{General formulas}
\label{general}

We work in the nonrecoil limit of an infinitely heavy 
nucleus and ignore the mixing of $2P_{1/2}$ and $2P_{3/2}$
states due to the hyperfine splitting (this mixing is otherwise
described in Sec.~III~C of Ref.~\cite{Pa1996muonic}).
Under these assumptions, the 
relativistic magnetic dipole interaction of the
nuclear magnetic moment and an electron in is
given by the Hamiltonian
\begin{equation}
\label{Hhfs}
H_{\rm hfs} 
= -e \, \vec{\alpha} \cdot \vec{A}_{\rm hfs}(\vec{r}) 
= |e| \, \vec{\alpha} \cdot \vec{A}_{\rm hfs}(\vec{r}) \,,
\end{equation}
where the vector potential reads
\begin{equation}
\label{Ahfs}
\vec{A}_{\rm hfs}(\vec{r}) = \frac{1}{4 \pi}\, \frac{\vec{\mu} \times \vec{r}}{r^3} \,,
\end{equation}
so that
\begin{equation}
H_{\rm hfs} 
= \frac{|e| }{4 \pi} \, \frac{\vec{\alpha} \cdot ( \vec{\mu} \times \vec{r} )}{r^3} 
= \frac{|e| }{4 \pi} \, \frac{\vec{\mu} \cdot ( \vec{r} \times \vec{\alpha} )}{r^3} \,.
\end{equation}
Here, $\vec{\mu}$ denotes the operator of the nuclear magnetic moment. 
In this paper, we will use the convention of labelling the relativistic
operators by indices with lower-case letters and nonrelativistic Hamilton operators by 
indices with upper-case symbols.
For future reference, we give the magnetic field
corresponding to the vector potential (\ref{Ahfs}),
\begin{equation}
\label{Bhfs}
\vec{B} = \vec{\nabla} \times \vec{A}_{\rm hfs} =
\frac23\, \vec{\mu}\, \delta^3(r) +
\frac{3 (\vec{\mu} \, \cdot \, \hat{\vec{r}}) \, \hat{\vec{r}} -
\vec{\mu} }{4 \pi \, r^3}\,.
\end{equation}
The operator $H_{\rm hfs}$ acts in the space of the coupled electron-nucleus states
\begin{equation}
| F M_F I j \rangle = 
\sum_{M\mu} C^{F M_F}_{I M j \mu} \, | I M \rangle \, | j \mu \rangle \,,
\end{equation}
where  $I$ and $M$ is the nuclear spin and its projection, 
$j$ and $\mu$ is the total electron angular
momentum and its projections, and $F$ and $M_F$ is the total momentum of the
system and its projection (we denote the momentum projection by $\mu$ 
instead of $m$ in order to differentiate it from the electron mass $m$). 
With the help of the Wigner--Eckhart theorem,
the expectation value of $H_{\rm hfs}$ on the coupled wave functions can be
reduced to a matrix element evaluated on the electronic wave functions only,
\begin{align}
\label{EF}
& 
\langle F M_F I j | H_{\rm hfs} | F M_F I j \rangle 
\\[2ex]
& = \frac{|e| m}{4\pi} \,
\frac{|\vec\mu|}{I} \, \left(2  \xi_e(j) \right) \,
\left<  F M_F I j \left| \vec{I} \cdot \vec{j} \right|  F M_F I j \right>
\nonumber\\[2ex]
& = \frac{|e| m}{4\pi} \,
\frac{|\vec\mu|}{I} \, \xi_e(j) \,
\left[ F(F+1) - I(I+1) - j(j+1) \right] 
\nonumber\\[2ex]
& = \alpha \, \frac{g_N}{2} \, \frac{m}{m_p} \, 
 \xi_e(j) \,
\left[ F(F+1) - I(I+1) - j(j+1) \right] \,.
\nonumber\
\end{align}
We have used $|e|^2 = 4\pi\alpha$ and 
$|\vec\mu| = g_N I |e|/(2 m_p)$, where $g_N$ is
the nuclear $g$ factor and $|e|/(2 m_p)$ is the nuclear magneton.
The quantity $ \xi_e(j)$ depends on the electronic state only.
Let $|j \mu \rangle$  denote the electronic state with 
total angular momentum $j$ and angular momentum projection $\mu$
(we here suppress the orbital angular momentum in our notation
for the 
electronic state). Then, with the index $q$ denoting the 
vector component in the spherical basis, we have
\begin{align}
\label{gammae}
 \xi_e(j) = & \;
\frac{\left< j \, \mu^{\prime} \left| 
\frac{\displaystyle [\vec{r}\times \vec{\alpha}]_q}{ \displaystyle m r^3} \right| 
j \, \mu \right>}%
{2 \left< j \mu^{\prime} \left| j_q \right| j \mu \right>} =
\frac{\left< j \, \mu \left| 
\frac{\displaystyle [\vec{r}\times \vec{\alpha}]_0}{\displaystyle m r^3} \right| 
j \, \mu \right>}%
{2 \left< j \, \mu
\left|  j _0 \right| j \, \mu \right>} 
\nonumber\\[2ex]
=& \; \frac{1}{2 \mu} \, 
\left< j \, \mu \left| \frac{[\vec{r}\times \vec{\alpha}]_0}{m r^3} 
\right| j \, \mu \right>
\nonumber\\[2ex]
=& \; \left< j \, \tfrac12 \left| \frac{[\vec{r}\times \vec{\alpha}]_0}{m r^3} 
\right| j \, \tfrac12 \right> \,.
\end{align}
As evident from the second line in the above equation, $\xi_e(j)$
does not depend on the actual value of the 
momentum projection $\mu$ of the total 
angular momentum of the electron, 
as its dependence cancels between the numerator and
denomenator. In a similar way, the nuclear variables can be factorized out in
evaluations of various corrections to the HFS, reducing the
problem in hand to the evaluation of an expectation value of an operator on
electronic state with a definite angular momentum projection. In 
practical calculations, we always assume the angular 
momentum projection of the reference electron state to be $\mu  = \tfrac{1}{2}$,
as in the third line of Eq.~\eqref{gammae}.

In the nonrelativistic limit, the magnetic dipole interaction 
describing the HFS consists of three terms,
\begin{subequations}
\label{master}
\begin{align}
H_{\rm HFS} =& \frac{|e| m}{4 \pi} \, \vec{\mu} \cdot \vec{h} =
\frac{|e| m}{4 \pi} \,
\vec{\mu} \cdot ( \vec{h}_S + \vec{h}_D + \vec{h}_L ) \,,
\\[2ex]
\label{hS} 
\vec{h}_S =& \; 
\frac{4 \pi}{3 m^2} \, \vec{\sigma} \, \delta^3(r) \,,
\\[2ex]
\label{hD} \vec{h}_D =& \; 
\frac{ 3 \,\, \hat{r} (\vec{\sigma} \cdot \hat{r}) \, \, 
- \vec{\sigma} }{2 \, m^2 \, r^3} \,,
\\[2ex]
\label{hL}
\vec{h}_L =& \; 
\frac{\vec{L}}{m^2 \, r^3}\,.
\end{align}
\end{subequations}
Analogously to the relativistic case, the nuclear degrees of freedom in the
expectation value of the operator $H_{\rm HFS}$ are factorized out, and the
problem is reduced to an evaluation of the matrix element of the purely
electronic operator
\begin{equation}
\label{h0}
h_0 = \frac{4}{3 m^2} \, \sigma_0 \, \delta^3(r) +
\frac{ 3 \,\, (\vec{\sigma} \cdot \hat{r}) \, \, \hat{r}_0 
- \sigma_0 }{2 \pi m^2 r^3} 
+ \frac{L_0}{\pi \, m^2 \, r^3} \,,
\end{equation}
where $\hat{r}$ is the unity vector $\vec r/r$. 
The nonrelativistic limit of $ \xi_e(j)$ is 
\begin{align}
\xi^{\rm NR}_e(j) = & \; \left< j \tfrac12 |h_0|j \tfrac12 \right>=
\frac{\kappa}{|\kappa|} \, \frac{(Z\alpha)^3 m}{n^3 (2 \kappa + 1) (\kappa^2 - 
\tfrac14)} \,,
\end{align}
where $\kappa = (-1)^{j-l+1/2} \,
(2 j+ 1)$ is the Dirac angular quantum number.
We have defined $\xi$ so that it has dimension of mass (energy) and so that 
its normalization reproduces the characteristic $4/3$ prefactor for the 
Fermi splitting of $S$ states.
For $nP_{1/2}$ and $nP_{3/2}$ states, we have, respectively,
\begin{subequations}
\label{gammaNR}
\begin{align}
\label{gammaNR12}
\xi^{\rm NR}_e(\tfrac12) \equiv & \;
\xi^{\rm NR}_e(2P_{1/2}) =
\frac{4}{9} \, \frac{(Z\alpha)^3 m}{n^3}\,,
\\
\label{gammaNR32}
\xi^{\rm NR}_e(\tfrac32) = & \;
\xi^{\rm NR}_e(2P_{3/2}) =
\frac{4}{45} \, \frac{(Z\alpha)^3 m}{n^3}\,.
\end{align}
\end{subequations}

Various corrections to the HFS can be conveniently expressed in terms of
multiplicative corrections to the quantity $\xi^{\rm NR}_e(j)$,
\begin{equation}
\label{defgamma}
\xi^{\rm NR}_{\rm e}(j) \to
\xi^{\rm NR}_{\rm e}(j) \,
\left[ 1 + \delta \xi_e(j) \right] \,.
\end{equation}
The corresponding corrections to the position of the HFS sublevels will then be
\begin{equation}
\label{defEHFS}
\delta E_{\rm HFS} = E_F\, \delta \xi_e(j) \,,
\end{equation}
where $E_F$ (the Fermi energy) is the nonrelativistic limit of
Eq.~(\ref{EF}). 
In order to keep our notations concise, we define the normalization factor
\begin{equation}
{\cal N} = \frac{1}{\langle j\tfrac 12|h_{\rm 0}|j\tfrac12 \rangle} = 
\frac{1}{\xi^{\rm NR}_e(j)} \,,
\end{equation}
which will be extensively used throughout the paper. 
Another reason for our choice of the normalization of 
$h_0$ is that we can use this operator as a perturbation 
Hamiltonian for the ordinary nonrelativistic self-energy 
(with the correct physical dimension of mass/energy),
in order to evaluate the relative correction to the Fermi
energy, provided we use a reference state with 
angular momentum projection $\mu = 1/2$.

%
%
\section{Foldy--Wouthuysen Transformation}
\label{fw}

The Foldy--Wouthuysen transformation \cite{FoWu1950} is a convenient tool for
obtaining the nonrelativistic expansion of the Dirac Hamiltonian in
external fields. The idea is to construct such a unitary transformation of the
original Hamiltonian that the transformed Hamiltonian does not couple the upper and the
lower components of the Dirac wave function to a specified order in $Z\alpha$. In
our case, we choose the starting Hamiltonian $H_t$ to be the sum of the Dirac
Hamiltonian $H_{\rm rel}$,
\begin{equation}
\label{Hrel}
H_{\rm rel} =  
\vec{\alpha} \cdot \vec{p} + \beta \, m - \frac{Z\alpha}{r} \,.
\end{equation}
and the relativistic HFS interaction operator,
\begin{equation}
H_t = H_{\rm rel} + H_{\rm hfs}  \,.
\end{equation}
For the purpose of the present investigation we construct the
Foldy--Wouthuysen transformation $U$ that decouples the upper and the
lower components of the Dirac wave function up to order $(Z\alpha)^4$ for
contributions to the energy and up to order $(Z\alpha)^3$ for contributions
proportinal to the magnetic moment. 
Because the general paradigm of the 
Foldy--Wouthuysen transformation has been extensively
discussed in the literature for a large class of potentials 
\cite{Zw1961,Pa2005}, we skip details of the derivation
and just indicate the results.
However, and this is an important point of the current paper, 
we must keep in mind that the Foldy--Wouthuysen transformation 
reads
\begin{equation}
\label{SFW}
U = \exp(\ii S) \,,\qquad
S = -\ii \, \beta \, {\rm Odd}(H_t) \,,
\end{equation}
where ${\rm Odd}(H_t)$ represents the matrix 
of the odd components of $H_t$ in $4 \times 4$ spinor space,
when the $4\times 4$ matrix is broken up in $2\times 2$ 
sub-matrices~\cite{FoWu1950,BjDr1966}.
Because $H_t$ contains $H_{\rm hfs}$, 
the Foldy--Wouthuysen transformation changes as compared to an
ordinary Lamb-shift calculation~\cite{Je1996}.
The transformed Hamiltonian $H'_t$ is
\begin{subequations}
\begin{equation}
\label{Htrafo}
H'_t = U H_t U^{-1} = H_{\rm FW} + H_{\rm HFS}\,,
\end{equation}
where the first part $H_{\rm FW}$ does not depend on the nuclear moment.
The second part is just the nonrelativistic HFS operator (\ref{master}),
which we reproduce here on the basis of the Foldy--Wouthuysen transformation
of the total relativistic Hamiltonian $H_t$. 
The operator $H_{\rm FW}$ is a $4\times 4$ matrix in spinor space,
\begin{align}
\label{HFW}
H_{\rm FW} =& \; \beta \left(m + \frac{\vec p^{\,2}}{2m}\right) 
- \frac{Z\alpha}{r} 
\nonumber\\[2ex]
& \; - \beta \frac{\vec p^{\,4}}{8 m^3} 
+ \frac{\pi Z\alpha}{2 m^2} \delta^3(r) 
+ \frac{Z\alpha}{4 m^2 r^3} \vec\Sigma\cdot\vec L \,.
\end{align}
Here, the $\vec\Sigma = \left( \begin{array}{cc} 
\vec\sigma & 0 \\
0 & \vec\sigma \end{array} \right)$
are the $4 \times 4$ generalizations
of the $2 \times 2$ Pauli matrices $\vec\sigma$.
For the upper components of the
wave function, $H_{\rm FW}$ can be replaced by the 
$2 \times 2$ matrix
\begin{equation}
\label{Hupper}
H_{\rm FW} \to m + \frac{\vec p^{\,2}}{2m}
- \frac{Z\alpha}{r} 
- \frac{\vec p^{\,4}}{8 m^3} 
+ \frac{\pi Z\alpha}{2 m^2} \delta^3(r) 
+ \frac{Z\alpha}{4 m^2 r^3} \vec\sigma\cdot\vec L \,,
\end{equation}
where we identify the second and third term 
as the nonrelativistic Schr\"{o}dinger Hamiltonian $H_{\rm NR}$.
By contrast, for the lower components, 
the applicable $2 \times 2$ matrix is [up to order 
$(Z\alpha)^2$]
\begin{equation}
\label{Hlower}
H_{\rm FW} \to -m - \frac{\vec p^{\,2}}{2m} - \frac{Z\alpha}{r} \,.
\end{equation}
\end{subequations}
However, the transformation of the Hamiltonian is not the only 
effect of the Foldy--Wouthuysen transformation.
In order to see that we also have to transform the transition current
of the electron, we remember that the characteristic 
integrand of a self-energy calculation is~\cite{Mo1974a,Pa1993}
\begin{equation}
{\cal M} = \left< \psi \left| 
\alpha^i \exp({\rm i} \vec k \cdot \vec r) 
\frac{1}{H_t - E_t + \omega}
\alpha^i \exp(-{\rm i} \vec k \cdot \vec r) 
\right| \psi \right> \,,
\end{equation}
where $(\omega, \vec k)$ is the four-momentum of the 
virtual photon, and $E_t$ is the total energy (corresponding 
to the Hamiltonian $H_t$) of the relativistic reference state $\psi$.
In order to achieve a nonrelativistic expansion, we perform
the Foldy--Wouthuysen transformation and write $\cal M$ as
\begin{equation}
{\cal M} = \left< U \psi \left| 
J^i 
\frac{1}{U (H_t - E_t + \omega) U^{-1}}
(J^*)^i 
\right| U \psi \right> \,.
\end{equation}
Here, $| U \psi \rangle$ is the nonrelativistic 
eigenket corresponding to the Schr\"{o}diger--Pauli 
eigenstate (plus relativistic corrections and HFS-induced
wave-function corrections), and we see how the 
transformed Hamiltonian $H'_t$ in the denominator 
is obtained [comparing to Eq.~\eqref{Htrafo}].
The transformed current $J^i$, which reads 
\begin{equation}
\label{j}
J^i = U \vec\alpha \exp({\rm i} \vec k \cdot \vec r) U^{-1} = 
\vec\alpha \exp({\rm i} \vec k \cdot \vec r) + \frac{\vec p}{m} + \dots,
\end{equation}
contains the dipole current $\vec p/m$ and higher-order 
terms. These higher-order terms, in the 
absence of the HFS interaction, are listed in Ref.~\cite{Je1996}.
In the presence of the HFS interaction, we
find, however, an {\em additional} contribution to the current,
due to the replacement $\vec{p} \to \vec{p} - e \, \vec{A}_{\rm hfs}$
for the momentum of the electron in the presence of the 
hfs vector potential,
\begin{equation}
\label{deltaj}
\frac{\vec p}{m} \to \frac{\vec p}{m} + 
\left( \frac{|e| m}{4 \pi} |\vec\mu| \right) 
\frac{\delta \vec j}{m} \,, \qquad
\delta \vec j = \frac{\hat \mu \times \vec r }{m r^3}\,,
\end{equation}
where $\hat \mu$ is the unit vector $\vec\mu/|\vec \mu|$.
The contribution induced by $\delta \vec j$ vanishes if radiative 
corrections are evaluated for the reference $S$ states,
but not for $P$ states which are investigated here.
In Eq.~\eqref{deltaj}, we define the prefactor 
multiplying $\delta \vec j$ in a way to be consistent with the normalization
of the operator $\vec h$ in Eq.~\eqref{master}. As discussed in the
previous section, the nuclear degrees of freedom can be effectively separated
out by assuming that the magnetic moment of the 
nucleus is pointing into the $z$ direction. 
In this case, the correction to the current takes the following form
in Cartesian coordinates,
\begin{equation}
\label{deltaj0}
\delta \vec j \to 
\delta \vec j_0 = 
- \frac{y}{m r^3} \hat{e}_x 
+ \frac{x}{m r^3} \hat{e}_y  \,.
\end{equation}
We note that, stricktly speaking, both the operator 
$\vec p/m$ on the right-hand side of Eq.~\eqref{j}
as well as the operator on the right-hand side of 
Eq.~\eqref{deltaj} should carry a $\beta$-matrix~\cite{Je1996}. However, it
can be replaced by unity when the current is applied 
to the upper components of the wave function, 
which are the only nonvanishing ones in the 
nonrelativistic approximation.

%
%
\section{Logarithmic Term}
\label{lterm}

In this section, 
we present a concise derivation of 
the logarithmic part of the self-energy contribution of order
$\alpha^2(Z\alpha)^2E_F$ to the HFS of $P$ states. 
To this end, we consider the perturbation of the nonrelativistic 
self-energy of the bound electron by the 
nonrelativistic hyperfine interaction $H_{\rm HFS}$.
Let $H_{\rm NR}$ be the Schr\"{o}dinger Hamiltonian 
and $H_T = H_{\rm NR} + H_{\rm HFS}$ denote the total nonrelativistic 
Hamiltonian, whose the reference-state eigenvalue will be denoted by $E_T$ and 
the corresponding eigenfunction, by $|\phi_T\rangle$. The 
nonrelativistic self-energy correction of the state $\phi_T$ is
\begin{align}
\delta E =& \; \frac{2 \alpha}{3 \pi}
\int_0^\epsilon d\omega \, \omega \, 
\left< \phi_T \left| \frac{p^i}{m} \, 
\frac{1}{E_T - (H_T  + \omega)} 
\frac{p^i}{m} \right| \phi_T \right> \,,
\end{align}
where $\epsilon$ is a non-covariant frequency cutoff
for the virtual photon. We use the expansion
\begin{equation}
\frac{1}{E_T - (H_T + \omega)} =
-\frac{1}{\omega} \, \left( 1 + \frac{E_T - H_T}{\omega} + 
{\cal O}\left( \frac{1}{\omega^2} \right) \right) 
\end{equation}
which is valid in the domain $\omega \in 
\left( (Z\alpha)^2 m, \epsilon \right)$ relevant to the calculation 
of the logarithm (note that in natural units, $\epsilon$ has dimension
of energy, or, equivalently, mass).
The logarithmic part of the correction is generated by the integral

\begin{align}
\delta E_{\rm log} =& \; \frac{2 \alpha}{3 \pi}
\int_{(Z\alpha)^2 m}^{\epsilon} d\omega \, \frac{1}{\omega} \, 
\left< \phi_T \left| \frac{p^i}{m} \, (H_T - E_T) \, 
\frac{p^i}{m} \right| \phi_T \right>
\\[2ex]
=& \; \frac{2 \alpha}{3 \pi}
\ln\left[ \frac{\epsilon}{(Z\alpha)^2 m} \right] \, 
\left< \phi_T \left| \frac{p^i}{m} \, (H_T - E_T) \, 
\frac{p^i}{m} \right| \phi_T \right>
\\[2ex]
=& \; \frac{\alpha}{3 \pi m^2}
\ln\left[ \frac{\epsilon}{(Z\alpha)^2 m} \right] \, 
\left< \phi_T \left| \left[ p^i, \, \left[ H_T , \, p^i \right] \right] 
\right| \phi_T \right> \,.
\end{align}
The parameter $\epsilon$ cancels at the
end of the calculation, when the high-energy part is added. 
For the determination of the logarithmic contribution it sufficient just to 
replace $\epsilon$ by the 
electron mass $m$.  To identify the correction of first order in $H_{\rm HFS}$, we expand
the reference-state wave function as
\begin{align}
| \phi_T \rangle = | \phi \rangle + 
\left( \frac{1}{E_{\rm NR} - H_{\rm NR}} \right)' H_{\rm HFS} | \phi \rangle \,,
\end{align}
where the prime denotes the reduced Green function.
The first-order perturbative correction of $\delta E_{\rm log}$  is
\begin{align}
& \delta E_{\rm log} \sim \frac{\alpha}{3 \pi m^2}
\ln\left[ (Z\alpha)^{-2} \right] \, 
\Biggl\{
\left< \phi \left| \left[ p^i, \, \left[ H_{\rm HFS} , \, p^i \right] \right] 
\right| \phi \right>
\nonumber\\
& + 
2\,\left< \phi \left| \left[ p^i, \, \left[ H_{\rm NR} , \, p^i \right] \right]
\left( \frac{1}{E_{\rm NR} - H_{\rm NR}} \right)' H_{\rm HFS} \right| \phi \right>
\Biggr\} \,.
\end{align}
For $P$ states, the second term in brackets vanishes because
$ \left[ p^i, \, \left[ H_{\rm NR} , \, p^i \right] \right]$ is proportional
to a Dirac $\delta$ function. Therefore, 
\begin{align}
\label{logterm}
\left. \delta \xi_e(j) \right|_{\rm log} =& \;
\frac{\alpha {\cal N}}{3 \pi m^2}
\ln[(Z\alpha)^{-2}] \, 
\left< j \tfrac{1}{2} \left| \left[ p^i, \, \left[ h_0 , \, p^i \right] \right] 
\right| j \tfrac{1}{2} \right> %
\nonumber\\[2ex]
 =& \;
\frac{\alpha {\cal N}}{3 \pi m^2}
\ln[(Z\alpha)^{-2}] \, 
 \left< j \tfrac{1}{2} \left| 
\vec\nabla^2 h_0  \right| j \tfrac{1}{2} \right> 
 \,.
\end{align}
Here, $|j \tfrac12 \rangle$ is the
Schr\"{o}dinger--Pauli eigenstate 
$| n P_j \rangle$ with angular momentum projection $\tfrac12$.
For $P$ states, we obtain
\begin{equation}
\label{nabla2H}
\left< j \tfrac{1}{2} \left| \frac{\vec\nabla^2 h_0}{m^2} 
\right| j \tfrac{1}{2} \right>=
-\frac83 \, \frac{n^2-1}{n^2} \, \frac{(Z\alpha)^5 m}{n^3} \,
\delta_{j, \tfrac12}\,.
\end{equation}
The Kronecker symbol in the above equation implies that 
the matrix element vanishes for $P_{3/2}$ states. The final result for the
logarithmic part of the correction is
\begin{equation} \label{eq000}
\left. \delta \xi_e(j) \right|_{\rm log} = 
-2 \, \frac{n^2-1}{n^2} \, 
\frac{\alpha}{\pi} \, (Z\alpha)^2 \, \ln[ (Z\alpha)^{-2} ] \,
\delta_{j, \tfrac12}.
\end{equation}

So, the self-energy correction to the HFS of 
$P$ states can be conveniently expressed as
\begin{equation}
\label{deltaSE}
\delta \xi_e(j) = \frac{\alpha}{\pi} 
\left[ a_{00} + (Z \alpha)^2 
\left\{ a_{21} \, \ln[(Z\alpha)^{-2}] + a_{20} \right\}+\ldots \right]\,,
\end{equation}
where $\ldots$ denote the higher-order terms.
As usual, the first index of $a_{ij}$ counts the power of
$Z\alpha$, and the second one indicates the power of the logarithm.

The result (\ref{eq000}) confirms the estimates of the logarithmic coefficient 
$a_{21}$ derived for the $2P_j$ states in Ref.~\cite{YeJe2009}
on the basis of an analysis of numerical data. Specifically, the values of 
$a_{21}(2P_{1/2})=-1.5$ and $a_{21}(2P_{3/2})=0.0$ were reported in 
that work, in
full agreement with the analytical result of Eq.~(\ref{eq000}).

%
%
\section{High--Energy Part}
\label{hep}

In this section we derive the part of the self-energy correction to order 
$\alpha(Z\alpha)^2E_F$ induced by virtual photons of high frequency, which is
referred to as the high-energy part. It can be obtained from the Dirac-Coulomb
Hamiltonian modified by the presence of the free-electron form factors $F_1$
and $F_2$ (for a derivation see, e.g., Chap.~7 of~\cite{ItZu1980}), 
\begin{align}
\label{HDm}
H_{\rm rad} =& \; \vec{\alpha} \cdot
\left[\vec{p} -{\mathrm e} \, F_1(\vec{\nabla}^2) \, \vec{A}\right]
+ \beta\,m + F_1(\vec{\nabla}^2) \, V
\nonumber\\
& + F_2(\vec{\nabla}^2) \, \frac{e}{2\,m} \, \left({\mathrm i}\,
\vec{\gamma} \cdot \vec{E} - \beta \, \vec{\Sigma} \cdot \vec{B}
\right)\,,
\end{align}
where $V = -Z\alpha/r$ is the Coulomb potential.
The form factors present in this Hamiltonian lead to various radiative corrections
to the HFS, when $\vec{A}$ and $\vec{B}$ are replaced by the vector potential
and the magnetic field corresponding to the hyperfine interaction,
respectively. For $S$ states, this procedure is described in detail in Ref.~\cite{JeYe2006}.

We find that the Hamiltonian (\ref{HDm}) induces six contributions of order
$\alpha(Z\alpha)^2E_F$, 
\begin{equation}
\delta \xi^{\rm H}_{\rm e}(j) =
\sum_{i=1}^6 {\cal C}_i \,,
\end{equation}
each of which will be addressed in turn in the following.

The first correction ${\cal C}_1$ 
is induced by a term with $F_2(0)$ in Eq.~(\ref{HDm}), namely
\begin{equation}
\label{deltaHhfs} 
-F_2(0) \, \frac{e}{2 m} \, \beta \,
\vec{\Sigma} \cdot \vec{B} = 
\frac{\alpha}{2 \pi} \, 
\left[ \frac{|e| m}{4 \pi} \, \beta \, 
\vec\mu \cdot (\vec h_s + \vec h_d) \right]\,.
\end{equation}
Here $\beta$ is the Dirac $\gamma^0$ matrix in the Dirac representation,  
$F_2(0) = \tfrac{\alpha}{2\pi}$, and 
the vectors $\vec h_s$ and $\vec h_d$ are the $4 \times 4$ generalizations
of $\vec h_S$ and $\vec h_D$, respectively,
\begin{subequations}
\label{HSHD} 
\begin{align}
\label{hs} 
\vec h_s =& \; 
\frac{4 \pi}{3 m^2} \, \vec{\Sigma} \, \delta^3(r) \,,
\\[2ex]
\label{hd} 
\vec h_d =& \; 
\frac{ 3 \,\, \hat{r} (\vec{\Sigma} \cdot \hat{r}) \, \, 
- \vec{\Sigma} }{2 \, m^2 \, r^3} \,.
\end{align}
\end{subequations}
The corresponding correction ${\cal C}_1$ is
\begin{equation}
\label{C1pre}
{\cal C}_1 = \frac{\alpha {\cal N}}{2\pi} \,
\left< j \tfrac12 \left| \beta \, (h_{s,0} + h_{d,0}) 
\right| j \tfrac12 \right>_R
 \,,
\end{equation}
where $h_{s,0}$ and $h_{d,0}$ are 
the $z$ components of the 
Hamiltonian operators defined in Eq.~\eqref{HSHD}.
We note that ${\cal C}_1$ contains the leading form-factor contribution of order
$\alpha$. To derive the next-order $\alpha(Z\alpha)^2$ correction, one has to
evaluate the matrix element with the relativistic (Dirac) 
wave functions (which have to be expanded in powers of $Z\alpha$ beforehand in order 
to escape divergences due to higher-order terms). By the index $R$, we denote
the matrix elements evaluated on the relativistic wave functions.  

The second correction (${\cal C}_2$) is an
$F_2'$ correction to the effective potential
(\ref{deltaHhfs}), i.e.,
\begin{align}
& -F'_2(0) \, \frac{e}{2m} \,
\beta \, \vec{\nabla}^2 \, \vec{\sigma}\cdot\vec{B} =
\nonumber\\
& \qquad \qquad \frac{\alpha}{12 \pi} \,
\left[ \frac{|e| m}{4 \pi} \, 
\beta \, 
\vec\mu \cdot 
\left\{ \vec{\nabla}^2 \, (\vec h_s + \vec h_d) \right\} \right]\,.
\end{align}
where we have used $F'_2(0) = \frac{\alpha}{12 \pi}$.
Up to the order $\alpha \, (Z\alpha)^2 \, E_F$, we 
can approximate the relativistic operators $\vec h_s$
and $\vec h_d$ by their nonrelativistic counterparts
$\vec h_S$ and $\vec h_D$, replace the $\beta$ matrix by unity,
and write
\begin{equation}
\label{C2pre}
{\cal C}_2 = \frac{\alpha {\cal N}}{12 \pi} \,
\left< j \tfrac12 \left|
\vec{\nabla}^2 \, \left( h_{S,0} + h_{D,0} \right) 
\right| j \tfrac12 \right> \,,
\end{equation}
to be evaluated on the nonrelativistic wave functions,

The third correction ${\cal C}_3$ is given by the term with $F_1'$
in Eq.~(\ref{HDm}), namely
\begin{align}
-{\mathrm e}\, F'_1(0) \, \vec{\nabla}^2 \vec{\alpha}\cdot\vec A = \frac{\alpha}{3\,\pi}\,
\left[ \ln\left( \frac{m}{2\,\epsilon} \right) + \frac{11}{24} \right]\,
\vec{\nabla}^2 H_{\rm hfs}  \,,
\nonumber
\end{align}
where $\epsilon$ is a noncovariant low-energy photon cut-off in the 
slope of the form factor $F_1$.
We can formulate ${\cal C}_3$ nonrelativistically,
\begin{equation}
{\cal C}_3 =
\frac{\alpha {\cal N}}{3\,\pi}\,
\left[ \ln\left( \frac{m}{2\,\epsilon} \right) + \frac{11}{24} \right]\,
\left< j \tfrac12 \left| \vec{\nabla}^2 h_0 
\right| j \tfrac12 \right> \,.
\end{equation}

The forth contribution to the high-energy part 
is a second-order perturbative correction induced by the
form-factor correction to the Coulomb potential $V$, 
\begin{align}
\frac{\alpha}{\pi} V_4 \equiv& \; \left[ F_1(\vec\nabla) - 1 \right] V 
\nonumber\\
=& \; \frac{\alpha}{3 \pi}\, (Z\alpha) \, \left[ \ln
\left(\frac{m}{2\,\epsilon}\right) + \frac{11}{24} \right] \,
\frac{\vec{\nabla}^2}{m^2} V + \dots\,.
\end{align}
The corresponding correction is
\begin{equation}
{\cal C}_4 = \frac{2 \alpha {\cal N}}{\pi}\,
\left<j \tfrac12 \left| 
V_4 \, \left( \frac{1}{E_{\rm NR} - {H_{\rm NR}}} \right)' \, 
h_0 \right| j \tfrac12 \right> \,,
\end{equation}
where we have done a Foldy--Wouthuysen transformation 
on the propagator denominator in order to obtain the 
Schr\"{o}dinger Hamiltonian $H_{\rm NR}$ and ignored 
higher-order (in $Z\alpha$) terms. Because $V_4$ is 
proportional to a Dirac $\delta$ function, the correction ${\cal C}_4$ 
vanishes for $P$ states.

The next contribution is the second-order perturbative correction
induced by the relativistic
hyperfine potential $H_{\rm hfs}$ as given in Eq.~(\ref{Hhfs}) and
the following term in Eq.~(\ref{HDm})
\begin{equation}
\frac{\alpha}{\pi} V_{56} \equiv 
F_2(0) \, \frac{e}{2\,m} \, {\mathrm i}\,
\vec{\gamma} \cdot \vec{E} =
- {\mathrm i}\, \frac{\alpha}{4 \pi m} \, 
\vec{\gamma} \cdot \vec\nabla V \,,
\end{equation}
where $\vec{E}$
is the electric field generated by the Coulomb potential $V$.
The total correction then is 
\begin{align}
\label{C56}
{\cal C}_5 + {\cal C}_6 
=& 
\; \frac{2 \alpha}{\pi}\,
\frac{ \left< \phi \left| 
V_{56} \, 
\left( \frac{1}{E_{\rm rel} - H_{\rm rel}} \right)' \, 
H_{\rm hfs} \right| \phi \right>_R }
{ \left< \phi \left| H_{\rm hfs} \right| \phi \right> }
\,.
\end{align}
It is conveniently splitted into two parts, ${\cal C}_5$ and $ {\cal C}_6$, as will be
discussed below. We note that the relativistic HFS interaction $H_{\rm hfs}$
couples the upper and the lower components of the wave function,
as does $V_{56}$, so that the second-order matrix element has to be evaluated
on the relativistic wave functions (which is indicated by the index $R$).

Let us consider the Foldy--Wouthuysen transformation of the 
numerator of the expression~\eqref{C56} very carefully. 
We write this numerator, employing a transformation 
${\cal U}$, as
\begin{equation}
\label{with_calU}
\left< \phi \left| {\cal U} V_{56} {\cal U}^{-1}\, 
\left( \frac{1}{{\cal U} (E_{\rm rel} - H_{\rm rel}) 
{\cal U}^{-1}} \right)' \, 
{\cal U} H_{\rm hfs} {\cal U}^{-1} \right| \phi \right>
\end{equation}
where ${\cal U}$, in contrast to $U$, is the 
Foldy--Wouthuysen transformation that diagonalizes the 
plain Dirac Hamiltonian~\eqref{Hrel} without
the HFS interaction (the state $|\phi\rangle$ after the 
transformation is just the nonrelativistic 
Schr\"{o}dinger--Pauli eigenstate). 
In particular $\cal U$ is obtained from Eq.~\eqref{SFW}
by the replacement $H_t \to H_{\rm rel}$.
It has been shown in Ref.~\cite{Je1996} that the 
Foldy--Wouthuysen transformation, when applied to a
``third-party'' operator---such as the 
electron transition current operator 
$\vec\alpha \, \exp(\ii \vec{k} \cdot \vec{r})$---leaves 
the leading term intact, and gives rise to higher-order 
corrections. We obtain in the case of $V_{56}$,
\begin{equation}
\label{V56}
{\cal U} V_{56} {\cal U}^{-1} = V_{56} + \frac{1}{8 m^2} \vec\nabla^2 V +
\frac{\alpha}{4 \pi} \, \frac{ Z\alpha }{m^2 \, r^3} \,
\vec{\sigma} \cdot \vec{L} + \dots 
\end{equation}
Note that when we add the term $\frac{1}{8 m^2} \vec\nabla^2 V$,
multiplied by the prefactor $\alpha/\pi$, to the potential $V_4$,
then we obtain the effective one-loop Lamb shift potential,
\begin{equation}
\frac{\alpha}{\pi}\, \left(V_4 + \frac{1}{8 m^2} \vec\nabla^2 V \right) =
\frac{\alpha}{3 \pi}\, (Z\alpha) \, 
\left[ \ln\left(\frac{m}{2\,\epsilon}\right) 
+ \frac{5}{6} \right] \, \frac{\vec{\nabla}^2}{m^2} V \,,
\end{equation}
which is useful when the entire formalism is applied to
$S$ states [see Eq.~(21) of Ref.~\cite{JeYe2006}].
Finally, and somewhat surprisingly, the transformation
${\cal U} \neq U$ applied to the relativistic 
hyperfine interaction
$H_{\rm hfs}$ leads to
\begin{equation}
\label{hfsHFS}
{\cal U} H_{\rm hfs} {\cal U}^{-1} = H_{\rm hfs} + H_{\rm HFS} + \dots \,.
\end{equation}
We retain the original relativistic HFS potential 
$H_{\rm hfs}$ as the 
leading term and obtain the full nonrelativistic
interaction $H_{\rm HFS}$ as an additional term, 
as well as higher-order terms which we
can ignore. The Hamiltonian in the propagator 
denominator is transformed with the help of
\begin{equation}
{\cal U} H_{\rm rel} {\cal U}^{-1} = H_{\rm FW} \,,
\end{equation}
where $H_{\rm FW}$ is the $4 \times 4$  Foldy--Wouthuysen Hamiltonian 
given in Eq.~\eqref{HFW}, which breaks up into the upper and 
lower components given in 
Eqs.~\eqref{Hupper} and~\eqref{Hlower}, respectively.
From the upper components and the third term on the 
right-hand side of~\eqref{V56}, we have
\begin{align}
{\cal C}_5 =& \; \frac{\alpha}{2 \pi}\,
\frac{ \left< \phi \left| 
\frac{ Z\alpha }{m^2 \, r^3} \,
\vec{\sigma} \cdot \vec{L} 
\left( \frac{1}{E_{\rm NR} -H_{\rm NR}} \right)' \, 
H_{\rm HFS} \right| \phi \right> }
{ \left< \phi \left| H_{\rm HFS} \right| \phi \right> }
\nonumber\\[2ex]
=& \; \frac{\alpha {\cal N}}{2 \pi}\,
\left< j \tfrac12 \left| 
\frac{ Z\alpha }{m^2 \, r^3} \,
\vec{\sigma} \cdot \vec{L} 
\left( \frac{1}{E_{\rm NR} -H_{\rm NR}} \right)' \, 
h_0 \right| j \tfrac12 \right> \,.
 \label{eq002}
\end{align}
The correction ${\cal C}_6$ is obtained when lower components 
of the Green function are selected. The bound-state energy can be 
approximated by $m$, and the energy of the virtual state as $-m$
[see Eq.~\eqref{Hlower}], and we obtain from the first term 
on the right-hand side of~\eqref{V56},
\begin{align}
{\cal C}_6 =& \; \frac{2 \alpha}{\pi}\,
\frac{ \left< \phi \left| V_{56} \,\, \frac{1}{2 m} \, 
H_{\rm hfs} \right| \phi \right>_R }
{ \left< \phi \left| H_{\rm HFS} \right| \phi \right> }
\\[2ex]
=& \; \frac{\alpha {\cal N}}{\pi}\,
\left( -\frac{1}{24} \, \delta_{j,1/2} +
\frac{1}{60} \, \delta_{j,3/2} \right)
 \left< j \tfrac12 \left| 
\frac{Z\alpha}{m^3 r^4} \right| j \tfrac12 \right>  \,.
 \label{eq003}
\end{align}
One might wonder what would have happened if we 
had done the Foldy--Wouthuysen transformation in
Eq.~\eqref{with_calU} with the full Foldy--Wouthuysen operator 
$U$, which also diagonalizes the HFS interaction.
In that case, we would not have obtained the term $H_{\rm hfs}$
on the right-hand side of~\eqref{hfsHFS}, but the 
transformation of the potential $V_{56}$ would have yielded an 
additional term, proportional to the nuclear magnetic moment.

In order to check the above derivation of the sum of the ${\cal C}_5$ and
${\cal C}_6$ corrections, we evaluate Eq.~(\ref{C56}) in a different way, by
using generalized virial relations~\cite{Sh1991,Sh2003} for the Dirac equation. 
First, we introduce the 
relativistic perturbed wave function $\delta \phi$ as 
\begin{align}
| \delta\phi \rangle = \left( \frac{1}{E_{\rm rel} - H_{\rm rel}} \right)' \, 
\frac{[\vec{r}\times\vec{\alpha}]_0}{m r^3}\, | j \tfrac12 \rangle\,.
\end{align}
We note that the operator $\frac{[\vec{r}\times\vec{\alpha}]_0}{m r^3}$ in the
above expression is the electronic part of $H_{\rm hfs}$ after the separation
of the nuclear degrees of freedom, see Eq.~(\ref{gammae}).
Performing the angular integration in Eq.~(\ref{C56}), we obtain (for an
arbitrary reference state)
\begin{align} \label{eq001}
{\cal C}_5+{\cal C}_6 &\ = -\frac{Z\alpha^2 {\cal N}}{2\pi m} 
\int_0^{\infty} dr\, [g(r)\, \delta f(r)+f(r)\, \delta g(r)]\,,
\end{align}
where $g$ and $f$ are the upper and the lower components of the (relativistic)
reference-state wave function, respectively, 
and $\delta g$ and $\delta f$ are those of the
(diagonal in $\kappa$ part of the) relativistic perturbed wave function $\delta \phi$. 
The perturbed wave-function components $\delta g$ and $\delta f$ are known in
closed analytical form \cite{Sh1991,Sh2003}. The radial integral
in Eq.~(\ref{eq001}) diverges for plain Dirac wave functions 
because of logarithmic singularities induced by 
higher-order terms in the $Z\alpha$
expansion of the integrand. One first has to expand the wave
functions in $Z\alpha$ and then perform the integration. The result
obtained in this way coincides with the one derived from Eqs.~(\ref{eq002})
and (\ref{eq003}).

We now summarize the high-energy corrections,
\begin{subequations}
\begin{align}
\label{C1}
{\cal C}_1 =& \frac{\alpha {\cal N}}{2\pi} \,
\left< j \tfrac12 \left| \beta \, 
(h_{s,0} + h_{d,0}) \right| j \tfrac12 \right>_R \,,
\\[2ex]
\label{C2}
{\cal C}_2 =& \frac{\alpha {\cal N}}{12 \pi} \,
 \left<  j \tfrac12 \left|
\vec{\nabla}^2 \, \left( h_{S,0} + h_{D,0} \right) 
\right|  j \tfrac12 \right> \,,
\\[2ex]
\label{C3}
{\cal C}_3 =&
\frac{\alpha  {\cal N}}{3\,\pi}\,
\left[ \ln\left( \frac{m}{2\,\epsilon} \right) + \frac{11}{24} \right]\,
\left< j \tfrac12 \left| \vec{\nabla}^2 h_0
\right| j \tfrac12 \right>\,,
\\[2ex]
\label{C5}
{\cal C}_5 =& \frac{\alpha {\cal N}}{2\,\pi}\,
\left< j\tfrac12 \left| 
 \frac{ Z\alpha }{m^2 \, r^3} \,
\vec{\sigma} \cdot \vec{L}  \, 
\left( \frac{1}{E_{\rm NR} - H_{\rm NR}} \right)' \, 
h_0 \right| j\tfrac12 \right> \,,
\\[2ex]
\label{C6}
{\cal C}_6 =& \frac{\alpha {\cal N}}{\pi}\,
\left( -\frac{1}{24} \, \delta_{j,1/2} +
\frac{1}{60} \, \delta_{j,3/2} \right)
 \left< j \tfrac12 \left| 
\frac{Z\alpha}{m^3 r^4} \right| j \tfrac12 \right>  \,.
\end{align}
\end{subequations}
An evaluation leads to the following results for $nP_{1/2}$ states
\begin{subequations}
\begin{align}
\label{C1P12}
{\cal C}_1(\tfrac12) =& 
\frac{\alpha}{\pi} \left[ \frac14 +
(Z\alpha)^2 \left( \frac{13}{48} + \frac{3}{8n} - \frac{7}{16 n^2} \right) \right] \,,
\\[2ex]
\label{C2P12}
{\cal C}_2(\tfrac12) =& 
\frac{\alpha}{\pi} (Z\alpha)^2 \frac{1 - n^2}{3 n^2} \,, 
\\[2ex]
\label{C3P12}
{\cal C}_3(\tfrac12) =& 
\frac{\alpha}{\pi} (Z\alpha)^2 \left\{
-\frac{n^2-1}{n^2} \left[ 2 \ln\left(\frac{m}{2\epsilon}\right) 
+ \frac{11}{12} \right] \right\} \,,
\\[2ex]
\label{C5P12}
{\cal C}_5(\tfrac12) =& 
\frac{\alpha}{\pi} (Z\alpha)^2 
\left( \frac{227}{180} + \frac{1}{2 n} - \frac{3}{5 \, n^2} \right) \,,
\\[2ex]
\label{C6P12}
{\cal C}_6(\tfrac12) =& 
- \frac{\alpha}{\pi} (Z\alpha)^2  \,
\frac{3 n^2 - 2}{20 \, n^2} \,,
\end{align}
\end{subequations}
and those for $nP_{3/2}$ states,
\begin{subequations}
\begin{align}
\label{C1P32}
{\cal C}_1(\tfrac32) =& \;
\frac{\alpha}{\pi} \left[ -\frac18 +
(Z\alpha)^2 \left( -\frac{119}{960} - \frac{3}{32 \, n} +
\frac{7}{20 n^2} \right) \right] \,,
\\[2ex]
\label{C2P32}
{\cal C}_2(\tfrac32) =& \;
\frac{\alpha}{\pi} (Z\alpha)^2  \,
\frac{5 (n^2 - 1)}{12 \, n^2} \,,
\\[2ex]
\label{C3P32}
{\cal C}_3(\tfrac32) =& \; 0\,,
\\[2ex]
\label{C5P32}
{\cal C}_5(\tfrac32) =& \;
\frac{\alpha}{\pi} (Z\alpha)^2  \,
\left( -\frac{227}{360} - \frac{1}{4 n} + \frac{3}{10 n^2} \right) \,,
\\[2ex]
\label{C6P32}
{\cal C}_6(\tfrac32) =& \;
\frac{\alpha}{\pi} (Z\alpha)^2  \,
\frac{3 n^2 - 2}{10 \, n^2} \,.
\end{align}
\end{subequations}
The net result for the high-energy part of the self-energy correction to the 
HFS of $P$ states is
\begin{align} \label{eq004}
\delta \xi^{\rm H}_e(\tfrac12) =& \;
\frac{\alpha}{\pi} \Biggl\{
\frac14 + (Z\alpha)^2 \left[ \frac{19}{144} + \frac{7}{8 \, n} +
\frac{5}{16 n^2} \right.
\nonumber\\
& \; \left. - 2 \frac{n^2-1}{n^2} \ln\left( \frac{m}{2 \epsilon} \right)
\right]  \Biggr\}\,,
\\[2ex]
\delta \xi^{\rm H}_e(\tfrac32) =& \;
\frac{\alpha}{\pi} \Biggl\{
-\frac18 + (Z\alpha)^2 \left[ -\frac{109}{2880} - \frac{11}{32 \, n} +
\frac{1}{30 n^2} \right] \Biggr\}.
 \label{eq005}
\end{align}
%

%
%
\section{Low--Energy Part}
\label{lep}

In this section we derive the part of the self-energy correction to order 
$\alpha(Z\alpha)^2E_F$ induced by virtual photons of low frequency, which is
referred to as the low-energy part in the following. 
In order to keep the notation concise, we suppress the 
indication of the reference state $| j \tfrac12 \rangle$ in the subsequent 
formulas; it is assumed that all matrix elements in the 
low-energy part are evaluated with this reference state.

The low-energy contribution is conveniently separated into four parts that can be
interpreted as corrections to the Hamiltonian,
to the reference-state wave function, 
to the reference-state energy, and to
the current.  The correction to the Hamiltonian is expanded first in $Z\alpha$,
then in the overlapping parameter $\epsilon$. The standard procedure (see, e.g., 
Refs.~\cite{Pa1993,JePa1996}) yields
\begin{align}
& \delta \xi^{\rm L}_H(j) 
= \frac{2 \alpha {\cal N}}{3 \pi} 
\int\limits_0^\epsilon {\rm d}\omega\, \omega 
\nonumber\\[1ex]
& \times 
\left< \frac{p^i}{m} \frac{1}{E_{\rm NR} - H_{\rm NR} - \omega} 
h_0 \frac{1}{E_{\rm NR} - H_{\rm NR} - \omega} \frac{p^i}{m} \right> 
\nonumber\\[1ex]
& \sim 
\frac{2 \alpha {\cal N}}{3 \pi m^2} 
\ln\!\left[ \frac{\epsilon}{(Z\alpha)^2 m} \right]
\left( \frac{1}{2} \left< [ p^i , [ h_0, p^i ]] \right> \!+\!
\left< p^2 h_0 \right> \right) 
\nonumber\\[1ex]
& \qquad + \frac{\alpha}{\pi} (Z\alpha)^2 \, \beta_H(j) \, .
\end{align}
Here, $\beta_H(j)$ is a Bethe-logarithm type correction
which needs to be evaluated numerically.
The correction to the wave function is expanded into 
logarithmic and nonlogarithmic parts as follows,
\begin{align}
& \delta \xi^{\rm L}_\psi(j) = 
\frac{4 \alpha {\cal N}}{3 \pi} 
\int\limits_0^\epsilon  {\rm d}\omega\, \omega 
\nonumber\\[1ex]
& \times 
\left< \frac{p^i}{m}
\frac{1}{E_{\rm NR} - H_{\rm NR} - \omega} \frac{p^i}{m}
\left( \frac{1}{E_{\rm NR} - H_{\rm NR}} \right)' h_0 \right>  
\nonumber\\[2ex]
& \sim 
\frac{2 \alpha {\cal N}}{3 \pi m^2} \,
\ln \! \left[ \frac{\epsilon}{(Z\alpha)^2 m} \right] \,
\biggl(
\left< \vec p^2 \right> \, \left< h_0 \right> -
\left< \vec p^2 h_0 \right> 
\nonumber\\[2ex]
& + 2 \left< [ p^i, [ (H_{\rm NR} - E_{\rm NR}), p^i]] 
\left( \frac{1}{E_{\rm NR} - H_{\rm NR}} \right)' h_0 \right>
\biggr) 
\nonumber\\[2ex]
& +\frac{\alpha}{\pi} (Z\alpha)^2 \,  \beta_\psi(j) \,,
\end{align}
where we have used commutation relations.
The correction to the energy is
\begin{align}
& \delta \xi^{\rm L}_E(j) =
- \frac{2 \alpha}{3 \pi} 
\int\limits_0^\epsilon {\rm d}\omega\, \omega 
\left< \frac{p^i}{m} \left( \frac{1}{E_{\rm NR} - H_{\rm NR} - \omega} \right)^2 
\frac{p^i}{m} \right> 
\nonumber\\[2ex]
& = - \frac{2 \alpha}{3 \pi m^2} \,
\ln \left[ \frac{\epsilon}{(Z\alpha)^2 m} \right] \,
\left< \vec p^{\,2} \right> + \frac{\alpha}{\pi} (Z\alpha)^2 \, \beta_E(j) \,.
\end{align}
The low-energy correction due to the nuclear-spin 
dependent current is    
\begin{align}
\delta \xi^{\rm L}_J(j) =& \;
\frac{4 \alpha {\cal N}}{3 \pi} \,
\int\limits_0^\epsilon {\rm d}\omega \, \omega \,
\left< \frac{p^i}{m} \frac{1}{E - H - \omega} 
\frac{\delta j^i_0}{m} \right> 
\nonumber\\[2ex]
=& \; \frac{\alpha}{\pi} (Z\alpha)^2 \,  \beta_J(j) \,,
\end{align}
where $\delta \vec j_0$ is defined in Eq.~\eqref{deltaj0}. 
Note that the correction to the current is ultraviolet finite
and therefore does not contribute to the logarithmic 
part of the correction.
The sum of all discussed corrections gives the 
low-energy part $\delta \xi^{\rm L}_e(j)$,
\begin{align}
\label{LEP}
& \delta \xi^{\rm L}_e(j) =
\delta \xi^{\rm L}_H(j) +
\delta \xi^{\rm L}_\psi(j) +
\delta \xi^{\rm L}_E(j) +
\delta \xi^{\rm L}_J(j) 
\nonumber\\[2ex]
& = \frac{\alpha {\cal N}}{3 \pi m^2} \,
\ln\!\left[ \frac{\epsilon}{(Z\alpha)^2 m} \right] \,
\biggl( \left< [ p^i, \, [  h_{\rm 0}, \, p^i ]] \right> 
\nonumber\\[2ex]
& + 2 \left< [ p^i, [ H_{\rm NR}, p^i]] \frac{1}{(E - H)'} h_0 \right> \biggr) 
+ \frac{\alpha}{\pi} (Z\alpha)^2 \, \beta(j) 
\nonumber\\[2ex]
& = -2\, (Z\alpha^2)\,\frac{n^2-1}{n^2}\,\ln\left( \frac{m}{2\epsilon}\right) 
\,\delta_{j,\tfrac12}
+ \frac{\alpha}{\pi} (Z\alpha)^2 \, \beta(j) \,,
\end{align}
where
\begin{equation}
\beta(j) = \beta_H(j) + \beta_\psi(j) + \beta_E(j) + \beta_J(j) \,.
\end{equation}
In Eq.~(\ref{LEP}), we took into account that
$\left< [ p^i, [ H_{\rm NR}, p^i]] \frac{1}{(E - H)'} h_0 \right>$ 
vanishes and the matrix element
$\left< [ p^i, \, [  h_{\rm 0}, \, p^i ]] \right>$ is evaluated in 
Eq.~\eqref{nabla2H}. It can be immediately seen that the sum $\delta \xi^{\rm L}_e(j)$
and $\delta \xi^{\rm H}_e(j)$ is free from $\epsilon$-dependent terms.

The results of our numerical evaluation for
the specific contributions for the $2P_{1/2}$ state are
\begin{subequations}
\begin{align}
\beta_H(2P_{1/2}) =& \; 0.33712 \,,
\\[2ex]
\beta_\psi(2P_{1/2}) =& \; 2.12732 \,,
\\[2ex]
\beta_E(2P_{1/2}) =& \; -0.12830\,,
\\[2ex]
\beta_J(2P_{1/2}) =& \; -0.52003\,,
\end{align}
\end{subequations}
and thus
\begin{equation}
\label{beta2P12}
\beta(2P_{1/2}) = 1.81611\,.
\end{equation}
For the $2P_{3/2}$ state, we have
\begin{subequations}
\begin{align}
\beta_H(2P_{3/2}) =& \; -0.32557\,,
\\[2ex]
\beta_\psi(2P_{3/2}) =& \; 2.12732\,,
\\[2ex]
\beta_E(2P_{3/2}) =& \; -0.12830 \,,
\\[2ex]
\beta_J(2P_{3/2}) =& \; -1.30008 \,,
\end{align}
\end{subequations}
and therefore
\begin{equation}
\label{beta2P32}
\beta(2P_{3/2}) = 0.37337\,.
\end{equation}
%

%
%
\section{Vacuum polarization}
\label{sec:vp}

The leading vacuum-polarization correction of order $\alpha(Z\alpha)^2E_F$
is induced by a matrix element of the radiatively corrected external magnetic
field (corresponding to a diagram with the hyperfine interaction 
inserted into the
vacuum-polarization loop). The corresponding correction is (see, e.g., 
Ref.~\cite{JeYe2006})
\begin{align}
\label{vp} 
& \delta  \xi_e^{\rm VP} =  
{\cal N}\, \frac{\kappa}{j(j+1)}\,
\, \frac{2\alpha}{3\pi}\,
\int_1^{\infty}dt\, \frac{\sqrt{t^2-1}}{t^2}\,
\left(1+\frac1{2t^2}\right)\,
\nonumber\\
& \quad \times \int_0^{\infty}dr\, ( 1+2m rt)\, e^{-2m rt}\,
g(r)\, f(r)\,,
\end{align}
where $g(r)$ and $f(r)$ are the upper and the lower components of the
reference-state wave function. To leading order in $\Za$, this expression
is evaluated for $P$ states to yield
\begin{align}
\label{vpres} 
 \delta  \xi_e^{\rm VP}(j) =
 \frac{\alpha}{\pi}\,(Z\alpha)^2\,\frac25\,\frac{n^2-1}{n^2}\,
\delta_{j,\tfrac12}.
\end{align}
Up to order $\alpha (Z\alpha)^2 E_F$ the vacuum-polarization 
correction thus vanishes for $P_{3/2}$ states.

\begin{table}[htb]
\caption{\label{table1} Higher-order remainder functions 
$g_{1/2}(Z\alpha)$ and $g_{3/2}(Z\alpha)$ for the $2P_{1/2}$ 
and $2P_{3/2}$ states, respectively, as obtained 
recently in Ref.~\cite{YeJe2009}. The value of $\alpha$
employed in the calculation is $\alpha^{-1} = 137.036$, and 
the numerical uncertainty of the all-order (in $Z\alpha$) 
calculation due to the finite number of integration
points is indicated in brackets.}
\begin{tabular}{c@{\hspace{0.4cm}}.@{\hspace{0.4cm}}.}
\hline 
\hline 
$Z$ & 
\multicolumn{1}{c}{$g_{1/2}(Z\alpha)$} & 
\multicolumn{1}{c}{$g_{3/2}(Z\alpha)$} \\
\hline 
1 &  3.437x410(86) &  0.126x09(18) \\
2 &  3.369x521(25) &  0.079x405(55) \\
3 &  3.300x616(14) &  0.032x176(39) \\
4 &  3.231x062(9)  & -0.015x499(29) \\
5 &  3.161x062(6)  & -0.063x528(21) \\
6 &  3.090x757(4)  & -0.111x933(16) \\
7 &  3.020x233(3)  & -0.160x663(12) \\
8 &  2.949x549(2)  & -0.209x698(9) \\
9 &  2.878x748(2)  & -0.259x028(7) \\
10&  2.807x853(2)  & -0.308x643(5) \\
\hline 
\hline 
\end{tabular}
\end{table}

\begin{figure}[thb]
\includegraphics[width=0.7\linewidth]{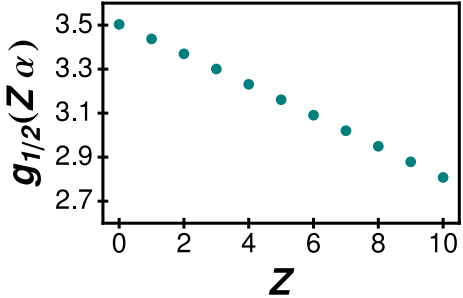}
\caption{\label{fig1} (Color online) We show the higher-order 
remainder function $g_{1/2}(Z\alpha)$ as a function 
of $Z$. Numerical values for $g_{1/2}(Z\alpha)$ 
are given in Table~\ref{table1}. The point at $Z=0$
is given by the coefficient $a_{20}(2P_{1/2})$.}
\end{figure}

\begin{figure}[thb]
\includegraphics[width=0.7\linewidth]{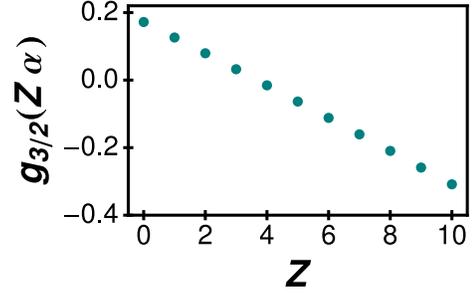}
\caption{\label{fig2} (Color online) Same as Fig.~\ref{fig1}, but for the 
$2P_{3/2}$ state.  The higher-order
remainder function $g_{3/2}(Z\alpha)$ is plotted as a function
of $Z$, with numerical values for $g_{3/2}(Z\alpha)$
given in Table~\ref{table1}. The point at $Z=0$
is $\lim_{Z\alpha \to 0} g_{3/2}(Z\alpha) = a_{20}(2P_{3/2})$.}
\end{figure}

%
%
\section{Final Results}
\label{final}

Summarizing our calculations, we conclude that the
QED correction to the HFS of $P$ states can be
cast into the form
\begin{align}
& \Delta E_{nP_j} = E_F(nP_j)\, \delta \xi_e(j) 
\\[2ex]
& = E_F(nP_j) \frac{\alpha}{\pi} 
\left[ a_{00} + (Z\alpha)^2 \, 
\left( a_{21} \, \ln[(Z\alpha)^{-2}] + a_{20} \right) 
\right] ,
\nonumber
\end{align}
which is valid up to order $\alpha (Z\alpha)^2 E_F$.
According to Eq.~\eqref{eq004}, 
the results for the leading coefficients are
\begin{equation}
a_{00}(nP_{1/2}) = \frac14 \,, \qquad
a_{00}(nP_{3/2}) = -\frac18 \,.
\end{equation}
The logarithmic part of the correction, as given by Eq.~(\ref{eq000}), is
\begin{equation}
a_{21}(nP_{1/2}) = -2 \frac{n^2 - 1}{n^2} \,, \qquad
a_{21}(nP_{3/2}) = 0 \,.
\end{equation}
In particular, there are no squared logarithmic terms.
As follows from Eqs.~(\ref{eq004}), (\ref{eq005}), (\ref{LEP}), and (\ref{vpres}),
the total nonlogarithmic contribution $a_{20}$ is the sum of 
a self-energy (SE) and a vacuum-polarization (VP) correction,
\begin{subequations}
\begin{equation}
a_{20} = a^{\rm SE}_{20} + a^{\rm VP}_{20} \,.
\end{equation}
The results read
\begin{align}
a^{\rm SE}_{20}(nP_{1/2}) = &\ 
\frac{19}{144} + \frac{7}{8 \, n} +
\frac{5}{16 n^2}  + 2 \frac{n^2-1}{n^2} \ln 2 
\nonumber \\[1ex]
& \ + \beta(nP_{1/2}) \,,
\\[2ex]
a^{\rm VP}_{20}(nP_{1/2}) = &\
\frac25\,\frac{n^2-1}{n^2} \,,
\\[2ex]
a^{\rm SE}_{20}(nP_{3/2}) = &\
 \frac{109}{2880} - \frac{11}{32 \, n} +
\frac{1}{30 n^2} + \beta(nP_{3/2}) \,,
\\[2ex]
a^{\rm VP}_{20}(nP_{3/2}) = &\ 0 \,.
\end{align}
\end{subequations}
Using the results for the $\beta$ terms in 
Eqs.~\eqref{beta2P12} and~\eqref{beta2P32},
we obtain, in particular,
\begin{subequations}
\begin{align}
a^{\rm SE}_{20}(2P_{1/2}) =& \; 3.50343 \,,
\\[2ex]
\label{a20_2P32}
a^{\rm SE}_{20}(2P_{3/2}) =& \; 0.17198 \,.
\end{align}
\end{subequations}
These results can be compared to numerical data at low $Z$
presented in Ref.~\cite{YeJe2009}.

Indeed, for low-$Z$ one-electron ions, a nonperturbative (in $Z\alpha$) 
calculation of the self-energy correction to the 
hyperfine splitting has recently been carried out~\cite{YeJe2009}. 
The numerical values for the self-energy correction $\Delta E_{nP_j}$
to the hfs find a natural representation as
\begin{align}
\label{remainder}
& \Delta E_{nP_j} = 
E_F(nP_j) 
\nonumber\\[2ex]
& \times 
\frac{\alpha}{\pi} 
\left[ a_{00} + (Z\alpha)^2 \, 
\left( a_{21} \, \ln[(Z\alpha)^{-2}] + g_j(Z\alpha)  \right) 
\right] ,
\nonumber
\end{align}
where $g_j(Z\alpha)$ is a remainder function which approaches
the $a_{20}$ coefficient for $Z\alpha \to 0$,
\begin{equation}
\lim_{Z\alpha \to 0} g_j(Z\alpha) = a_{20}(nP_j)\,.
\end{equation}
Numerical values for the remainder functions $g_j(Z\alpha)$ are given in 
Table~\ref{table1}. In Figs.~\ref{fig1} and~\ref{fig2},
we plot the higher-order remainder against the nuclear charge number $Z$.
The numerical data are consistent with the next higher-order term 
in the expansion of $\Delta E_{nP_j}$ being a correction of order
$\alpha (Z\alpha)^3$ (no logarithm).

%
%
\section{Conclusions}
\label{conclu}

The hyperfine structure of $P$ states is an interesting physical 
problem. In accurate measurements of the classical ($2P_j$--$2S$) Lamb shift 
in atomic hydrogen, both the hyperfine effects of $P$ as well as 
of $S$ states have to be carefully accounted for before a meaningful comparison
of theory and experiment can be made. In alkali-metal atoms, the hyperfine
structure of $P$ states is also of great experimental interest 
\cite{GeDeTa2003,WaEtAl2003,DaNa2006}.

In the present investigation, we analyze QED corrections
to the hyperfine splitting of $nP_{1/2}$ and $nP_{3/2}$ states
in hydrogenlike systems, up to order $\alpha (Z\alpha)^2 E_F$,
where $E_F$ is the Fermi splitting. Our calculation relies
on the separation of the electronic from the nuclear degrees of 
freedom as described in Sec.~\ref{general}, effectively reducing the 
problem to an electronic self-energy type calculation. 
The identification of the nonrelativistic 
degrees of freedom relevant to our investigation is 
accomplished by the Foldy--Wouthuysen transformation 
as described in Sec.~\ref{fw}.
A nuclear-spin dependent correction to the electronic 
transition current is identified [see Eq.~\eqref{deltaj}].
We show (see Sec.~\ref{lterm})
that squared logarithmic corrections of relative order
$\alpha (Z\alpha)^2 \ln^2[(Z\alpha)^{-2}] E_F$ are completely absent for
$P$ states, whereas for $P_{3/2}$, even the single
logarithmic term of relative order
$\alpha (Z\alpha)^2 \ln[(Z\alpha)^{-2}] E_F$ vanishes.
This finding is interesting in view of different conjectures
described in the literature~\cite{SaCh2006}.

In order to address
the nonlogarithmic correction of relative order
$\alpha (Z\alpha)^2 E_F$, we split the calculation into a 
high- and a low-energy part and match them via an intermediate
overlapping parameter $\epsilon$ that separates the scales
of high-energy and low-energy photons
(see Secs.~\ref{hep} and~\ref{lep}). This parameter is noncovariant 
but turns out to lead to a concise formulation of a problem 
which is otherwise rather involved.
The high-energy part is treated in Sec.~\ref{hep} and is 
seen to lead to form-factor type corrections.
For the low-energy part treated in Sec.~\ref{lep},
the correction to the electron's transition current 
induced by the hyperfine interaction is crucial.
This correction can be obtained via a Foldy--Wouthuysen 
transformation (Sec.~\ref{fw}).
The discussion of vacuum-polarization 
corrections (see Sec.~\ref{sec:vp}) and a brief summary
of the results obtained (Sec.~\ref{final}) conclude our
investigation.

We reemphasize once more that
the logarithmic $a_{21}$ coefficient vanishes 
for $P_{3/2}$ states [see Eq.~\eqref{deltaSE}], 
and the $a^{\rm SE}_{20}$ self-energy coefficient
[see Eq.~\eqref{a20_2P32}] also is numerically small for 
$2P_{3/2}$.  These two observations account for the numerically 
small results obtained in the all-order calculation~\cite{YeJe2009}
for the self-energy corrections to the HFS of this state.
Indeed, the QED self-energy corrections to the hyperfine splitting of $2P_{3/2}$ states
are surprisingly small at low $Z$.
This behaviour is naturally attributed to the less singular 
behaviour of the $P_{3/2}$ states at the origin in comparison to
that of the $P_{1/2}$ states. 

%
%
\section*{Acknowledgments}

U.D.J.~has been supported by the National Science Foundation (Grant
PHY--8555454) as well as by a Precision Measurement Grant from the National
Institute of Standards and Technology.  V.A.Y.~was supported by DFG (grant
No.~436 RUS 113/853/0-1) and acknowledges support from RFBR (grant
No.~06-02-04007) and the foundation ``Dynasty.''


\begin{thebibliography}{10}

\bibitem{MoPlSo1998}
P.~J. Mohr, G. Plunien, and G. Soff, Phys. Rep. {\bf 293},  227  (1998).

\bibitem{Mo1974a}
P.~J. Mohr, Ann. Phys. (N.Y.) {\bf 88},  26  (1974).

\bibitem{Mo1974b}
P.~J. Mohr, Ann. Phys. (N.Y.) {\bf 88},  52  (1974).

\bibitem{GuEtAl2005}
A. Gumberidze, T. St\"{o}hlker, D. Ban\'{a}s, K. Beckert, P. Beller, H.~F.
  Beyer, F. Bosch, S. Hagmann, C. Kozhuharov, D. Liesen, F. Nolden, X. Ma,
  P.~H. Mokler, M. Steck, D. Sierpowski, and S. Tashenov, Phys. Rev. Lett. {\bf
  94},  223001  (2005).

\bibitem{JePa1996}
U. Jentschura and K. Pachucki, Phys. Rev. A {\bf 54},  1853  (1996).

\bibitem{Pa1993}
K. Pachucki, Ann. Phys. (N.Y.) {\bf 226},  1  (1993).

\bibitem{FoWu1950}
L.~L. Foldy and S.~A. Wouthuysen, Phys. Rev. {\bf 78},  29  (1950).

\bibitem{ErYe1965ab}
G.~W. Erickson and D.~R. Yennie, Ann. Phys. (N.Y.) {\bf 35},  271, 447  (1965).

\bibitem{PaCzJeYe2005}
K. Pachucki, A. Czarnecki, U.~D. Jentschura, and V.~A. Yerokhin, Phys. Rev. A
  {\bf 72},  022108  (2005).

\bibitem{JePa2002}
U.~D. Jentschura and K. Pachucki, J. Phys. A {\bf 35},  1927  (2002).

\bibitem{JeMoSo1999}
U.~D. Jentschura, P.~J. Mohr, and G. Soff, Phys. Rev. Lett. {\bf 82},  53
  (1999).

\bibitem{NiEtAl2000}
M. Niering, R. Holzwarth, J. Reichert, P. Pokasov, {\relax Th}. Udem, M. Weitz,
  T.~W. H\"{a}nsch, P. Lemonde, G. Santarelli, M. Abgrall, P. Laurent, C.
  Salomon, and A. Clairon, Phys. Rev. Lett. {\bf 84},  5496  (2000).

\bibitem{FiEtAl2004}
M. Fischer, N. Kolachevsky, M. Zimmermann, R. Holzwarth, {\relax Th}. Udem,
  T.~W. H\"{a}nsch, M. Abgrall, J. Gr\"unert, I. Maksimovic, S. Bize, H.
  Marion, F. Pereira Dos~Santos, P. Lemonde, G. Santarelli, P. Laurent, A.
  Clairon, C. Salomon, M. Haas, U.~D. Jentschura, and C.~H. Keitel, Phys. Rev.
  Lett. {\bf 92},  230802  (2004).

\bibitem{JeMoSo2001pra}
U.~D. Jentschura, P.~J. Mohr, and G. Soff, Phys. Rev. A {\bf 63},  042512
  (2001).

\bibitem{YeInSh2004}
V.~A. Yerokhin, P. Indelicato, and V.~M. Shabaev, Phys. Rev. A {\bf 69},
  052503  (2004).

\bibitem{BlChSa1997reserve}
S.~A. Blundell, K.~T. Cheng, and J. Sapirstein, Phys. Rev. Lett. {\bf 78},
  4914  (1997).

\bibitem{SuPeSaScLiSo1998}
P. Sunnergren, H. Persson, S. Salomonson, S.~M. Schneider, I. Lindgren, and G.
  Soff, Phys. Rev. A {\bf 58},  1055  (1998).

\bibitem{YeSh2001hyp}
V.~A. Yerokhin and V.~M. Shabaev, Phys. Rev. A {\bf 64},  012506  (2001).

\bibitem{Pa1996}
K. Pachucki, Phys. Rev. A {\bf 54},  1994  (1996).

\bibitem{NiKi1997}
M. Nio and T. Kinoshita, Phys. Rev. D {\bf 55},  7267  (1997).

\bibitem{SaCh2006}
J. Sapirstein and K.~T. Cheng, Phys. Rev. A {\bf 74},  042513  (2006).

\bibitem{SaCh2008}
J. Sapirstein and K.~T. Cheng, Phys. Rev. A {\bf 78},  022515  (2008).

\bibitem{YeJe2009}
V.~A. Yerokhin and U.~D. Jentschura, Self-energy correction to the hyperfine
  splitting and the electron $g$ factor in hydrogen-like ions, Phys. Rev. A,
  submitted.

\bibitem{BrPa1968}
S.~J. Brodsky and R.~G. Parsons, Phys. Rev. {\bf 176},  423  (1968).

\bibitem{JeYe2006}
U.~D. Jentschura and V.~A. Yerokhin, Phys. Rev. A {\bf 73},  062503  (2006).

\bibitem{Pa1996muonic}
K. Pachucki, Phys. Rev. A {\bf 53},  2092  (1996).

\bibitem{Zw1961}
D. Zwanziger, Phys. Rev. {\bf 121},  1128  (1961).

\bibitem{Pa2005}
K. Pachucki, Phys. Rev. A {\bf 71},  012503  (2005).

\bibitem{BjDr1966}
J.~D. Bjorken and S.~D. Drell, {\em Relativistische Quantenmechanik}
  (Bibliographisches Institut, Mannheim, Wien, Z\"urich, 1966).

\bibitem{Je1996}
U.~D. Jentschura, {\em Master Thesis: The Lamb Shift in Hydrogenlike Systems,
  [in German: Theorie der Lamb--Verschiebung in wasserstoffartigen Systemen],}
  (University of Munich, 1996, unpublished (see e-print hep-ph/0305065)).

\bibitem{ItZu1980}
C. Itzykson and J.~B. Zuber, {\em \relax{Quantum Field Theory}} (McGraw-Hill,
  New York, 1980).

\bibitem{Sh1991}
V.~M. Shabaev, J. Phys. B {\bf 24},  4479  (1991).

\bibitem{Sh2003}
V.~M. Shabaev,  in {\em Precision Physics of Simple Atomic Systems -- Lecture
  Notes in Physics Vol.~627}, edited by S.~G. Karshenboim and V.~B. Smirnov
  (Springer, Berlin, 2003), pp.\ 97--113.

\bibitem{GeDeTa2003}
V. Gerginov, A. Derevianko, and C.~E. Tanner, Phys. Rev. Lett. {\bf 91},
  072501  (2003).

\bibitem{WaEtAl2003}
J. Walls, R. Ashby, J.~J. Clarke, B. Lu, and W.~A. van Wijngaarden, Eur. Phys.
  J. D {\bf 22},  159  (2003).

\bibitem{DaNa2006}
D. Das and V. Natarajan, J. Phys. B {\bf 39},  2013  (2006).

\end{thebibliography}
\end{document}